\documentclass[aps,amsfonts,pra,singlecolumn,superscriptaddress,showpacs]{revtex4}
\usepackage{epsfig,amsmath,amssymb,bm,epsf,graphics}
\usepackage{epsfig,amsmath,amssymb,bm,epsf,graphics}

\def\bra#1{\langle#1\vert}
\def\ket#1{\vert#1\rangle}
\def\ketbra#1{\vert#1\rangle\langle#1\vert}

\def\ipr#1#2{\langle#1\vert#2\rangle}

\def\Longarrow{\protect\@lra}
\def\@lra{\relbar\joinrel\relbar\joinrel\relbar\joinrel%
          \relbar\joinrel\rightarrow}
\def\wstate{{\rm W}}

\def\coe#1{{({\rm co}E_{#1})}}
\begin{document}
\title{Connections between relative entropy of entanglement 
       and geometric measure of entanglement}
 \author{Tzu-Chieh Wei}
 \affiliation{Department of Physics, 
 University of Illinois at Urbana-Champaign, 
 1110 West Green Street, Urbana, Illinois 61801-3080, U.S.A.}
 \author{Marie Ericsson}
 \affiliation{Department of Physics, 
  University of Illinois at Urbana-Champaign, 
  1110 West Green Street, Urbana, Illinois 61801-3080, U.S.A.}
 \affiliation{ 
    Institute for Quantum Computing, 
    Department of Physics, 
    University of Waterloo, 
    200 University Avenue West, 
    Ontario, Canada N2L 3G1}
 \author{Paul M. Goldbart}
 \affiliation{Department of Physics, 
  University of Illinois at Urbana-Champaign, 
  1110 West Green Street, Urbana, Illinois 61801-3080, U.S.A.}
 \author{William J. Munro}
 \affiliation{Hewlett-Packard Laboratories, Filton Road, Stoke Gifford,
          Bristol, BS34 SQ2, UK}
\date{July 3, 2004}
\begin{abstract}
As two of the most important entanglement measures---the entanglement of formation
and the entanglement of distillation---have so far been limited to bipartite settings, the study
of other entanglement measures for multipartite systems appears necessary.
Here, connections between two other entanglement measures---the relative entropy 
of entanglement and the geometric measure of entanglement---are investigated. 
It is found that for arbitrary pure states the latter gives rise to a lower 
bound on the former.  For certain pure states, some bipartite and some multipartite, 
this lower bound is saturated, and thus their relative entropy of entanglement 
can be found analytically in terms of their known geometric measure of entanglement. 
For certain mixed states, upper bounds on the relative entropy of entanglement are 
also established.  Numerical evidence strongly suggests that these upper bounds are 
tight, i.e., they are actually the relative entropy of entanglement. 
\end{abstract}
\maketitle
\section{Introduction}
Entanglement has been identified as a resource central to much of quantum 
information processing~\cite{NielsenChuang00}.  To date, progress 
in the quantification of entanglement for mixed states has resided 
primarily in the domain of bipartite systems~\cite{Horodecki01}.  
For multipartite systems in pure and mixed states the characterziation and quantification
of entanglement presents even greater challenges.  Even for multipartite 
pure states it is not clear whether there exists a finite minimal reversible
entanglement generating set (MREGS)~\cite{BennettPopescuRohrlichSmolinThapliyal}
and, if it exists, what the set is.  This complicates the task
of extending measures such as entanglement of distillation~\cite{BennettBernsteinPopescuSchumacher96} and
formation~\cite{BennettDiVincenzoSmolinWootters96,Wootters98}
to multipartite systems~\cite{PlenioVedral01}.  Moreover, the 
characterization of multipartite entanglement remains incomplete. 

On the other hand, quantifying multipartite entanglement via other 
measures, such as relative entropy of entanglement~\cite{VedralPlenioRippinKnight97,PlenioVedral01}, is
still a challenging task, even for {\it pure\/} states. One reason
for the difficulty is the absence, in general, of Schmidt 
decompositions for multipartite pure states~\cite{Peres95}. This
implies that for multipartite pure states the entropies of the reduced 
density matrices can differ, in contrast to bipartite pure states,
as the following example shows.  Consider a three-qubit pure state
$\ket{\psi}_{ABC}\equiv\alpha\ket{001}+\beta\ket{010}+\gamma\ket{100}$,
where $|\alpha|^2+|\beta|^2+|\gamma|^2=1$.
The reduced density matrices for parties A, B, and C are, respectively,
\begin{subequations}
\begin{eqnarray}
\rho_A&=&(|\alpha|^2+|\beta|^2)\ketbra{0}+|\gamma|^2\ketbra{1}, \\
\rho_B&=&(|\alpha|^2+|\gamma|^2)\ketbra{0}+|\beta|^2\ketbra{1},\\
\rho_C&=&(|\gamma|^2+|\beta|^2)\ketbra{0}+|\alpha|^2\ketbra{1},
\end{eqnarray}
\end{subequations}
which, in general, have different entropies.  Thus, for a multipartite pure 
state the entropy of the reduced density matrix does not give a consistent 
entanglement measure.  However, even in the case in which all parties 
have the identical entropy, e.g., 
$\alpha=\beta=\gamma=1/\sqrt{3}$~\cite{Note}, 
it is in general nontrivial to obtain the relative entropy of entanglement 
for the state.  More generally, for pure multipartite states, it is not yet 
known how to obtain their relative entropy of entanglement analytically.
The situation is even worse for {\it mixed\/} multipartite states.

Recently, a multipartite entanglement measure based on the geometry
of Hilbert space has been proposed~\cite{Shimony95,BarnumLinden01,WeiGoldbart03}.
For pure states, this geometric measure of entanglement depends
on the maximal overlap between the entangled state and unentangled
states, and is easy to compute numerically.  
The measure has been applied to several bipartite and
multipartite pure and mixed states~\cite{BarnumLinden01,WeiGoldbart03},
including two distinct multipartite bound entangled states~\cite{WeiAltepeterGoldbartMunro03}.  
In the present paper, we explore connections between this measure and 
the relative entropy of entanglement.  
For certain pure states, some bipartite and some multipartite, 
this lower bound is saturated, and thus their relative entropy of entanglement 
can be found analytically, in terms of their known geometric measure of entanglement. 
For certain mixed states, upper bounds on the relative entropy of entanglement are 
also established.  Numerical evidence strongly suggests that these upper bounds are 
tight, i.e., they are actually the relative entropy of entanglement.
These results, although not general enough to solve the problem
of calculating the relative entropy of entanglement for arbitrary
multipartite states, may offer some insight into, and serve as a testbed 
for, future analytic progress related to the relative entropy of entanglement.

The structure of the present paper is as follows.  In Sec.~\ref{sec:measures} 
we review the two entanglement measures considered in the paper: the relative 
entropy of entanglement and the geometric measure of entanglement.  In 
Sec.~\ref{sec:connection} we explore connections between the two, in both 
pure- and mixed-state settings.  Examples are provided in which bounds and 
exact values of the relative entropy of entanglement are obtained.  
In Sec.~\ref{sec:summary} we give some concluding remarks.

\section{Entanglement measures}
\noindent%
\label{sec:measures}%
In this section we briefly review the two measures considered
in the present paper: the relative entropy of entanglement and 
the geometric measure of entanglement.
\subsection{Relative entropy of entanglement}
\noindent
The relative entropy $S(\rho||\sigma)$ between two states 
$\rho$ and $\sigma$ is defined via
\begin{equation}
S(\rho||\sigma)\equiv {\rm Tr}\left(\rho\log_2\rho-\rho\log_2{\sigma}\right),
\end{equation}
which is evidently not symmetric under exchange of $\rho$ and $\sigma$, 
and is non-negative, i.e., $S(\rho||\sigma)\ge 0$.  The relative
entropy of entanglement (RE) for a mixed state $\rho$ is defined to be
the minimal relative entropy of $\rho$ over the set of separable mixed states~\cite{VedralPlenioRippinKnight97,VedralPlenio98}:
\begin{equation}
\label{eqn:ER}
E_{\rm R}(\rho)\equiv \min_{\sigma\in {\cal D}}S(\rho||\sigma)=\min_{\sigma\in {\cal D}}{\rm Tr}\left(\rho\log_2\rho-\rho\log_2\sigma\right),
\end{equation}
where ${\cal D}$ denotes the set of all separable states.  

In general, the task of finding the RE for arbitrary states $\rho$ 
involves a minimization over all separable states, and this renders
the computation of the RE very difficult.  For bipartite pure states, 
the RE is equal to entanglements of formation and of distillation.  
But, despite recent progress~\cite{Ishizaka03}, for mixed states---even 
in the simplest setting of two qubits---no analog of Wootters' 
formula~\cite{Wootters98} for the entanglement of formation has been found. 
Things are even worse in multipartite settings.  Even for pure states, 
there has not been a systematic method for computing relative entropies of 
entanglement. It is thus worthwhile seeking cases in which one can explicitly
obtain an expression for the RE.  A trivial case arises when there exists a 
Schmidt decomposition for a multipartite pure state: in this case,
the RE is the usual expression
\begin{equation}
-\sum_i \alpha_i^2 \log_2 \alpha_i^2\,,
\end{equation}
where the $\alpha_i$'s are Schmidt coefficients 
(with $\sum_i \alpha_i^2=1$).
We shall see that there exist cases in which the RE can be determined 
analytically, even though there is no Schmidt decomposition.

{We remark that an alternative definition of RE is to replace the set
of separable states by the set of postive partial transpose (PPT) states.
The RE thus defined, as well as its regularized version, gives a tighter
bound on distillable entanglement. There has been important progress in
calculating the RE (and its regularized version) with respect to PPT states
for certain bipartite mixed states; see Refs.~\cite{AudenaertEtAl} for
more detailed discussions. For multipartite settings one could
also use this definition, and define the set of states to optimize over
to be the set of states that are PPT with respect to all bipartite partitionings.
However, we shall use the first definition, i.e., optimization over the set
of completely separable states, throughout the discussion of the present paper.}

\subsection{Geometric measure of entanglement}
\noindent
We continue by briefly reviewing the formulation of this  
measure in both pure-state and mixed-state settings.   
Let us start with a multipartite system comprising $n$ parts, 
each of which can have a distinct Hilbert space.  Consider a 
general $n$-partite pure state (expanded in the local bases 
$\{|e_{p_i}^{(i)}\}$): 
\begin{equation}
|\psi\rangle=\sum_{p_1\cdots p_n}\chi_{p_1p_2\cdots p_n}
|e_{p_1}^{(1)}e_{p_2}^{(2)}\cdots e_{p_n}^{(n)}\rangle.
\end{equation}
As shown in Ref.~\cite{WeiGoldbart03}, the closest separable pure state, 
\begin{equation}
\ket{\phi}\equiv\mathop{\otimes}_{i=1}^n|\phi^{(i)}\rangle
=\mathop{\otimes}_{i=1}^{n}
\Big(\sum_{p_i}c_{p_i}^{(i)}\,|e_{p_i}^{(i)}\rangle\Big),
\end{equation}
satisfies the stationarity conditions
\begin{subequations}
\label{eqn:Eigen}
\begin{eqnarray}
\!\!\!\!\!\!\!\sum_{p_1\cdots\widehat{p_i}\cdots p_n}
\chi_{p_1p_2\cdots p_n}^*c_{p_1}^{(1)}\cdots\widehat{c_{p_i}^{(i)}}\cdots c_{p_n}^{(n)}=
\Lambda\,{c_{p_i}^{(i)}}^*, \\ 
\!\!\!\!\!\!\!\!\!\!\sum_{p_1\cdots\widehat{p_i}\cdots p_n}\chi_{p_1p_2\cdots p_n} {c_{p_1}^{(1)}}^*\cdots\widehat{{c_{p_i}^{(i)}}^*}\cdots {c_{p_n}^{(n)}}^*=
\Lambda\,c_{p_i}^{(i)}\,,
\end{eqnarray}
\end{subequations} 
in which the eigenvalues $\Lambda$ are associated with the Lagrange 
multiplier enforcing the constraint $\ipr{\phi}{\phi}\!=\!1$, and lie 
in $[-1,1]$, and the symbol \,\,$\widehat{}$\,\, denotes exclusion.  
Moreover, the spectrum of $\Lambda$'s can be interpreted as the cosine 
of the angle between $|\psi\rangle$ and $\ket{\phi}$; 
the largest, $\Lambda_{\max}$ (i.e. $\cos\theta_{\min}$ with the smallest angle $\theta_{\min}$), which we call the 
{\it entanglement eigenvalue\/}, corresponds to the closest 
separable state, and is the maximal overlap with unentangled states: 
\begin{equation}
\Lambda_{\max}(\ket{\psi})=\max_{\phi}|\ipr{\phi}{\psi}|,
\end{equation}
where $\ket{\phi}$ is an arbitrary separable pure state.  
In Ref.~\cite{WeiGoldbart03}, the particular form
$E_{\sin^2}\equiv 1-\Lambda^2_{\max}(\ket{\psi})=\sin^2\theta_{\min}$ was defined to be the
geometric measure of entanglement (GME) for any pure state $|\psi\rangle$. 
Here, we shall be concerned with the related quantity 
$E_{\log_2}(\psi)\equiv-2\log_2\Lambda_{\max}(\ket{\psi})$, 
which we shall show to be a lower bound on the relative entropy of 
entanglement for $\ket{\psi}$.  Although this quantity is not, 
as we shall see later, an entanglement monotone for mixed states, 
it is a good measure of {\it pure-state\/} entanglement.

Given the definition of entanglement for pure states just formulated, 
the extension to mixed states $\rho$ can be built upon pure states 
via the {\it convex hull\/} construction (indicated by ``co''), 
as was done for the entanglement of formation; see Ref.~\cite{Wootters98}. 
 The essence is a minimization 
over all decompositions $\rho=\sum_i p_i\,|\psi_i\rangle\langle\psi_i|$ 
into pure states: 
\begin{eqnarray}
\label{eqn:Emixed}
E(\rho)
\equiv
\coe{\rm pure}(\rho)
\equiv
{\min_{\{p_i,\psi_i\}}}
\sum\nolimits_i p_i \, 
E_{\rm pure}(|\psi_i\rangle).
\end{eqnarray}
This convex hull construction ensures that the measure gives zero 
for separable states; however, in general it also complicates the 
task of determining mixed-state entanglement.  

\smallskip
\noindent
{\it Illustrative examples\/}: 
We consider several examples involving symmetric states,
mostly restricting our attention to $n$-qubit systems. 
First, one can classify permutation-invariant pure states, as follows:
\begin{equation}
\label{eqn:Snk}
|S(n,k)\rangle\equiv \sqrt{\frac{k!(n-k)!}{n!}}
\sum_{\rm{\scriptstyle Permutations}}
{\rm P}|\underbrace{0\cdots0}_{k}\underbrace{1\cdots1}_{n-k}\rangle.
\end{equation}
As the amplitudes are all positive, one can assume that the closest 
separable (equivalently, Hartree) state is of the form
\begin{equation}
\label{eqn:separable}
|\phi\rangle=\big(\sqrt{p}\,|0\rangle+\sqrt{1-p}\,|1\rangle\big)^{\otimes n},
\end{equation} 
for which the maximal overlap (w.r.t.~$p$) gives the entanglement
eigenvalue for $|{\rm S}(n,k)\rangle$:
\begin{eqnarray}
\label{eqn:Lambda}
\Lambda_{\max}(n,k)=
\sqrt{\frac{n!}{k!(n\!-\!k)!}}
\left(\frac{k}{n}\right)^{\frac{k}{2}}
{\left(\frac{n-k}{n}\right)}^{\frac{n\!-\!k}{2}}.
\end{eqnarray}
More generally, for $n$ parties each a $(d+1)$-level
system, the state
\begin{equation}
\label{eqn:Snkk}
\ket{S(n;\{k\})}\equiv \sqrt{\frac{k_0!k_1!\cdots k_d!}{n!}}
\sum_{\rm{\scriptstyle Permutations}}{\rm P}\,
|\underbrace{0\ldots0}_{k_0}\,\underbrace{1\ldots1}_{k_1}\ldots
 \underbrace{d\ldots d\,}_{k_d}\,\rangle
\end{equation}
has the entanglement eigenvalue 
\begin{equation}
\Lambda_{\max}(n;\{k\})=\sqrt{\frac{n!}{\prod_i (k_i!)}}
\,\prod_{i=0}^{d}\left(\frac{k_i}{n}\right)^{\frac{k_i}{2}}.
\end{equation}
Now consider the totally antisymmetric state 
$\ket{{\rm Det}_n}$, defined via
\begin{equation}
\label{eqn:Detn}
\ket{{\rm Det}_n}\equiv
\frac{1}{\sqrt{n!}}
\sum_{i_1,\dots,i_n=1}^{n}
\epsilon_{i_1,\dots,i_n}\ket{i_1,\dots,i_n}.
\end{equation}
It has been shown~\cite{Bravyi02} that $\Lambda^2_{\max}=1/n!$.  
The generalization of the antisymmetric state to the $n=p\,d^p$-partite 
determinant state is via~\cite{Bravyi02}
\begin{eqnarray}
\phi(1)&=&(0,0,\dots,0,0), \nonumber \\
\phi(2)&=&(0,0,\dots,0,1), \nonumber \\
&\vdots& \nonumber \\
\phi(d^p-1)&=&(d-1,d-1,\dots,d-1,d-2), \nonumber \\
\phi(d^p)&=&(d-1,d-1,\dots,d-1,d-1),\nonumber
\end{eqnarray}
and 
\begin{equation}
\label{eqn:Detnd}
\ket{{\rm Det}_{n,d}}\equiv\frac{1}{\sqrt{(d^p!)}}\sum_{i_1,\dots,i_{d^p}}
\epsilon_{i_1,\dots,i_{d^p}}\ket{\phi(i_1),\dots,\phi(i_{d^p})}.
\end{equation}
In this case, it can be shown that $\Lambda^2_{\max}=1/(d^p)!$.

Although the above states were discussed in terms of the 
GME~\cite{WeiGoldbart03}, we shall, in the following section, 
show the rather surprising fact that the RE of these example states, 
is given by the corresponding expression: $-2 \log_2\Lambda_{\max}$.

\section{Connection between the two measures}
\label{sec:connection}
\noindent
In bipartite systems, due to the existence of Schmidt decompositions,
the relative entropy of entanglement of a pure state is simply 
the von Neumann entropy of its reduced density matrix.  However, 
for multipartite systems there is, in general, no such decomposition,  
and how to calculate the relative entropy of entanglement for an 
arbitrary pure state remains an open question.  
We now connect the relative entropy of entanglement to the
geometric measure of entanglement for arbitrary pure states
by giving a lower bound on the former in terms of the latter 
or, more specifically, via the entanglement eigenvalue.

\subsection{Pure states: 
lower bound on relative entropy of entanglement}
\noindent
Let us begin with the following theorem:

\smallskip
\noindent
{\it Theorem\/} 1. 
For any pure state $\ket{\psi}$ with entanglement eigenvalue
$\Lambda_{\max}(\psi)$ the quantity $-2\log_2\Lambda_{\max}(\psi)$ is a 
lower bound on the relative entropy of entanglement of $\ket{\psi}$, i.e.,
\begin{equation}
\label{eqn:2log}
E_{\rm R}(\ketbra{\psi})\ge -2\log_2\Lambda_{\max}(\psi).
\end{equation}

\smallskip\noindent
{\it Proof\/}: 
From the definition~(\ref{eqn:ER}) of the relative entropy of 
entanglement we have, for a pure state $\ket{\psi}$, 
\begin{equation}
E_{\rm R}(\ketbra{\psi})=
 \min_{\sigma\in {\cal D}}-\bra{\psi}\log_2\sigma\ket{\psi}=
-\max_{\sigma\in {\cal D}} \bra{\psi}\log_2\sigma\ket{\psi}.
\end{equation}
Using the concavity of the log function, we have 
\begin{equation}
\label{eqn:inequalityLog}
\bra{\psi}\log_2\sigma\ket{\psi}\le \log_2 (\bra{\psi}\sigma\ket{\psi})
\end{equation}
and, furthermore, 
\begin{equation}
\label{eqn:inequalityMax}
\max_{\sigma\in{\cal D}}
\bra{\psi}\log_2\sigma\ket{\psi}\le 
\max_{\sigma\in{\cal D}} 
\log_2 (\bra{\psi}\sigma\ket{\psi}),
\end{equation}
although the $\sigma$'s maximizing the left- and right-hand sides 
are not necessarily identical.  We then conclude that
\begin{equation}
E_{\rm R}(\ketbra{\psi})\ge 
-\max_{\sigma\in{\cal D}}
\log_2 (\bra{\psi}\sigma\ket{\psi}).
\end{equation}
As any $\sigma\in{\cal D}$ can be expanded as 
$\sigma=\sum_i p_i\ketbra{\phi_i}$, 
where ${\ket{\phi_i}}$'s are separable
pure states, one has
\begin{equation}
\bra{\psi}\sigma\ket{\psi}=
\sum_i p_i |\ipr{\phi_i}{\psi}|^2 \le 
\Lambda^{2}_{\max}({\psi}), 
\end{equation}
and hence we arrive at the sought result 
\begin{equation}
E_{\rm R}(\ketbra{\psi})\ge 
-2\log_2\Lambda_{\max}({\psi}).
\label{eq:theorem}
\end{equation}

\smallskip
\noindent
{We wish to point out that such an inequality was previously established
and exploited in Refs.~\cite{VidalEtAl}.}

{When does the inequality becomes an equality?
The demand that Eq.~(\ref{eqn:inequalityLog}) hold as an equality
implies that  $\sigma$ (un-normalized) can be decomposed into either 
(a)
\begin{subequations}
\begin{equation}
\label{eqn:case1}
\sigma=\sum_{i} \ketbra{i},
\end{equation}
where $\{\ket{i}\}$ are mutually orthogonal
but {\it not\/} orthogonal to $\ket{\psi}$,
or (b)
\begin{equation}
\label{eqn:case2}
\sigma= \ketbra{\psi}+\tau^\perp,
\end{equation}
\end{subequations}
where $\tau^\perp$ (either pure or mixed) is orthogonal to $\psi$, i.e., $\bra{\psi}\tau^\perp\ket{\psi}=0$.
However, the separable $\sigma$ that has either property is not
necessarily the one that maximizes both sides of the inequality~(\ref{eqn:inequalityMax}), 
unless $\ket{\psi}$ (and hence $\sigma$) has high symmetry. 
} 
On the other hand, a corollary arises from Thereom 1 which says that for any multipartite pure state $\ket{\psi}$,
if one can find a separable mixed state $\sigma$ such that 
$S(\rho||\sigma)\vert_{\rho=\ketbra{\psi}}=
-2\log_2\Lambda_{\max}\big(\ket{\psi}\big)$ 
then $E_{\rm R}=-2\log_2\Lambda_{\max}\big(\ket{\psi}\big)$. 
This result follows directly from the fact that when the 
lower bound on $E_{\rm R}$ given in Eq.~(\ref{eqn:2log}) equals an 
upper bound, the relative entropy of entanglement is immediate. 
In all the examples we shall consider for which this lower bound 
is saturated, it turns out that 
\begin{equation}
\label{eqn:sigma}
\sigma^*\equiv\sum_i p_i\, \ketbra{\phi_i}
\end{equation}
is a closest separable mixed state, in which $\{\ket{\phi_i}\}$ are 
separable pure states closest to $\ket{\psi}$. 
(The distribution $p_i$ is uniform, and 
can be either discrete or continuous, and $\{\ket{\phi_i}\}$ are not necessarily
mutually orthogonal.) 

We now examine several illustrative states in the light of the above 
corollary, thus obtaining $E_{\rm R}$ for each of them.  We begin with the 
permutation-invariant states $\ket{S(n,k)}$ of Eq.~(\ref{eqn:Snk}), 
for which $\Lambda_{\max}$ was given in Eq.~(\ref{eqn:Lambda}). 
The above theorem guarantees that 
$E_{\rm R}\big(\ket{S(n,k)}\big)\ge 
-2\log_2 \Lambda_{\max}(n,k)$. 
To find an {\it upper\/} bound we construct a separable mixed state 
\begin{subequations}
\begin{eqnarray}
\sigma^*
&\equiv& 
\int \frac{d\phi}{2\pi}\ketbra{\xi(\phi)},
\\
\ket{\xi(\phi)}
&\equiv& 
\left(\sqrt{p}\ket{0}+
e^{i\phi}\sqrt{1-p}\ket{1}\right)^{\otimes n}, 
\end{eqnarray}
\end{subequations}
with $p$ chosen to maximize 
$||\ipr{\xi}{S(n,k)}||=\sqrt{C_k^n \,p^k(1-p)^{n-k}}$, which gives $p=k/n$. 
Direct evaluation then gives 
\begin{equation}
\label{eqn:sigmastar}
\sigma^*=\sum_{k=0}^n C^n_k p^{k}(1-p)^{(n\!-\!k)}\ketbra{S(n,k)},
\end{equation}
and
$S(\rho||\sigma)=-2\log_2\Lambda_{\max}(n,k)$, where $\rho=\ketbra{S(n,k)}$
and $\Lambda_{\max}(n,k)$ is given in Eq.~(\ref{eqn:Lambda}).  
The upper and lower bounds on $E_{\rm R}$ coincide, 
and hence we have that  
\begin{equation}
\label{eqn:rhoSnk}
E_{\rm R}\big(\ket{S(n,k)}\big)=-2\log_2 \Lambda_{\max}(n,k).
\end{equation}
The closest separable
mixed state $\sigma^*$
belongs to the case (b), i.e.,~Eq.~(\ref{eqn:case2}).
Similar equalities can be established for the generalized 
permutation-invariant $n$-party $(d+1)$-dit states 
$\ket{S(n,\{k\})}$ of Eq.~(\ref{eqn:Snkk}). 
We remark that the entanglements of the symmetric states 
$\ket{S(n,k)}$ (which are also known as {\it Dicke\/} states) have 
been analyzed via other approaches; see Ref.~\cite{Stockton}.  

For our next example we consider the totally anti-symmetric states 
$\ket{{\rm Det}_n}$ of Eq.~(\ref{eqn:Detn}).  It was shown in 
Ref.~\cite{Bravyi02} that for these states $\Lambda^2_{\max}=1/n!$, 
and hence it is straightforward to see that each of the $n!$ basis 
states $\ket{i_1,\dots,i_n}$ is a closest separable pure state.  
Thus, one can construct a separable mixed state from these
separable pure states [cf.~Eq.~(\ref{eqn:sigma})]:
\begin{equation}
\sigma_1\equiv\frac{1}{n!}\sum_{i_1,\dots,i_n}
\ketbra{i_1,\dots,i_n}.
\end{equation}
Then, by direct calculation one gets 
$S(\rho_{{\rm Det}_n}||\sigma_1)=\log_2(n!)$, 
which is identical to 
$-2\log_2\Lambda_{\max}$, 
as mentioned above.  As in our previous examples, upper and lower 
bounds on $E_{\rm R}$ coincide, and hence we have that 
$E_{\rm R}(\ket{{\rm Det}_n})=\log_2(n!)$. The closest separable
mixed state $\sigma_1$ 
belongs to the case (a), i.e.,~Eq.~(\ref{eqn:case1}).
Similarly, for the generalized determinant state~(\ref{eqn:Detnd}) 
one can show that $E_{\rm R}=\log_2(d^p!)$.

We now focus our attention on three-qubit settings. Of these, 
the states $\ket{S(3,0)}=\ket{000}$ and $\ket{S(3,3)}=\ket{111}$ 
are not entangled and are, respectively, the components of the 
the 3-GHZ state:
$\ket{{\rm GHZ}}\equiv
\big(\ket{000}+\ket{111})/\sqrt{2}$. Although the GHZ state is not of
the form $\ket{S(n,k)}$, it has $\Lambda_{\max}=1/\sqrt{2}$,
and two of its closest separable
pure states are $\ket{000}$ and $\ket{111}$~\cite{WeiGoldbart03}. 
From these one can construct a separable mixed state
\begin{eqnarray}
\sigma_2&=&\frac{1}{2}\big(\ketbra{000}+\ketbra{111}\big), 
\end{eqnarray}
From the discussion given after Eq.~(\ref{eq:theorem}), one concludes
that $E_{\rm R}({\rm GHZ})=-2 \log_{2} \Lambda_{\max}=1$ 
and that $\sigma_2$ is one of
the closest separable mixed states to $\ket{\rm GHZ}$. This closest separable
mixed state $\sigma_2$ 
belongs to the case (a), i.e.,~Eq.~(\ref{eqn:case1}). With some rewriting,
it can also be classified as case (b), i.e.,
\begin{equation}
\sigma_2=\frac{1}{2}\ketbra{\rm GHZ}+\frac{1}{2}\ketbra{\rm GHZ^-},
\end{equation}
where $\ket{{\rm GHZ}}\equiv
\big(\ket{000}-\ket{111})/\sqrt{2}$.

The states 
\begin{subequations}
\begin{eqnarray}
|{\rm W}\rangle
&\equiv&
\ket{{S}(3,2)}=
\big(\ket{001}+\ket{010}+\ket{100}\big)/\sqrt{3},
\\
\ket{\tilde{\rm W}}
&\equiv& 
\ket{{S}(3,1)}=
\big(\ket{110}+\ket{101}+\ket{011}\big)/\sqrt{3},
\end{eqnarray}
\end{subequations}
are equally entangled, and have 
$\Lambda_{\max}=2/3$~\cite{WeiGoldbart03}. 
Again, from the discussion after Eq.~(\ref{eq:theorem}) we have 
$E_{\rm R}=\log_{2}(9/4)$, and one of the closest separable mixed 
states to the W state can be constructed from 
\begin{eqnarray}
\sigma_{3}&\equiv&
\int\frac{d\phi}{2\pi}\ketbra{\psi(\phi)},
\quad{\rm with}
\\
\ket{\psi(\phi)}&\equiv&
\big(\sqrt{2/3}\ket{0}+
e^{i\phi}\sqrt{1/3}\ket{1}\big)^{\otimes 3},
\end{eqnarray}
which gives the result 
\begin{eqnarray}
\sigma_{3}=
\frac{4}{9}\ketbra{{\rm W}}+\frac{2}{9}\ketbra{\tilde{\rm W}}
+
\frac{8}{27}\ketbra{000}+\frac{1}{27}\ketbra{111}.
\end{eqnarray}
We remark that the mixed state $\sigma_{3}$ is not the only closest 
separable mixed state to the W state;
the following state $\sigma_{4}$ is another example 
(as would be any mixture of $\sigma_{3}$ and $\sigma_{4}$):
\begin{subequations}
\begin{eqnarray}
\sigma_4
\equiv\frac{1}{3}\sum_{k=0}^{2}
\ketbra{\psi(2\pi k/3)}
=\frac{4}{9}\ketbra{{\rm W}}+\frac{2}{9}\ketbra{\tilde{\rm W}}+
\frac{1}{3}\ketbra{\xi},
\end{eqnarray}
\end{subequations}
where $3\ket{\xi}\equiv
{2\sqrt{2}}\ket{000}+\ket{111}$.  These closest separable mixed states
of W state
belong to the case (b), i.e.,~Eq.~(\ref{eqn:case2}).

\begin{figure}
\centerline{\psfig{figure=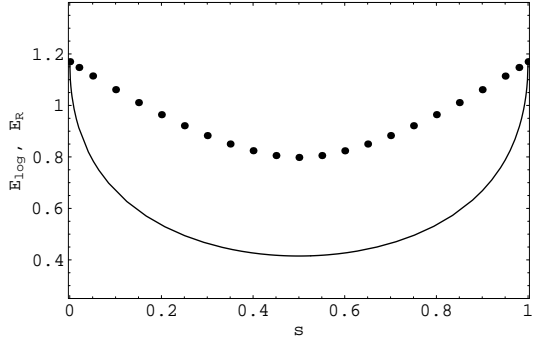,width=7cm,height=4.5cm,angle=0}}
\caption{The solid curve represents $E_{\log_{2}}(s)$ of the pure state
$\sqrt{s}\,|{\rm W}\rangle+\sqrt{1-s}\,|\tilde{\rm W}\rangle$ vs.~$s$. 
The dots are corresponding relative entropies of entanglement obtained 
numerically.}
\label{fig:ElogWW}
\end{figure}
Having obtained RE for ${\rm W}$ and $\tilde{\rm W}$, it is interesting
to examine the RE of the following superposition of the two:  
$\ket{\wstate\tilde{\rm W}(s)}
\equiv \sqrt{s}\,
\ket{\wstate}+\sqrt{1-s}\,\ket{\tilde{\rm W}}$. 
We  have not been able to find an analytical result for RE, 
but we can compare the analytical expression for 
$-2\log_2\Lambda_{\max}(\wstate\tilde{\rm W}(s))$ 
with the numerical evaluation of 
$E_{\rm R}(\wstate\tilde{\rm W}(s))$, 
and we do this in Fig.~\ref{fig:ElogWW}.  As we see in this figure, 
the qualitative behavior of the two functions is similar, but  
$-2\log_2\Lambda_{\max}$ and $E_{\rm R}$ only coincide at the two 
end-points, $s=0$ and $s=1$.

\subsection{Mixed states: upper bound on relative entropy of entanglement}
\noindent
In Ref.~\cite{WeiGoldbart03} the procedure was given to find the geometric 
measure of entanglement, $E_{\sin^2}$, for the mixed state comprising
symmetric states:
\begin{equation}
\label{eqn:mixSnk}
\rho(\{p\})=\sum_k p_k\,\ketbra{S(n,k)}.
\end{equation}
Here, we focus instead on the quantity $E_{\log_{2}}$, but the basic 
procedure is the same.  The first step is to find the entanglement 
eigenvalue 
$\Lambda_{n}(\{q\})$ 
for the pure state
\begin{equation}
\sum_k \sqrt{q_k}\,\ket{S(n,k)},
\end{equation}
thus arriving at the quantity
\begin{equation}
\label{eqn:epsilon}
{\cal E}(\{q\})\equiv -2 \log_2 \Lambda_{n}(\{q\}).
\end{equation}
Then the quantity $E_{\log_{2}}$ for the mixed state~(\ref{eqn:mixSnk}) 
is actually the convex hull of the expression~(\ref{eqn:epsilon}): 
\begin{equation}
E_{\log_2}\left(\rho(\{p\})\right)={\rm co}\,{\cal E}(\{p\}).
\end{equation}

This prompts us to ask the question:
Can we find RE for the mixture of $\ket{S(n,k)}$ in Eq.~(\ref{eqn:mixSnk})? 
To answer it, we shall first construct an upper bound to RE, and then compare 
this bound with the numerically evaluated RE.  To accomplish the first step, 
bearing in mind the fact that any separable mixed state will yield an upper 
bound, we consider the state formed by mixing the separable pure states 
$\ket{\xi(\theta,\phi)}$ [cf.~Eq.~(\ref{eqn:sigmastar})]: 
\begin{eqnarray}
\label{eqn:sigmatheta}
\sigma(\theta)=\int \frac{d\phi}{2\pi} \ketbra{\xi(\theta,\phi)}=\sum_{k=0}^n C^n_k\cos^{2k}\theta\sin^{2(n\!-\!k)}\theta\ketbra{S(n,k)},
\end{eqnarray}
where
\begin{equation}
\label{eqn:xi}
\ket{\xi(\theta,\phi)}\equiv 
\left(\cos\theta \ket{0}+ e^{i\phi}\sin\theta\ket{1}\right)^{\otimes n}.
\end{equation}
We then minimize the relative entropy between $\rho(\{p\})$ and $\sigma(\theta)$,
\begin{equation}
S\left(\rho(\{p\})\vert\vert\sigma(\theta)\right)=\sum_k p_k 
\log\frac{p_k}{C^n_k\cos^{2k}\theta\sin^{2(n\!-\!k)}\theta},
\end{equation}
with respect to $\theta$, obtaining the stationarity condition 
\begin{equation}
\label{eqn:theta}
\tan^2\theta\equiv \frac{\sum_k p_k\,(n-k)}{\sum p_k \,k}.
\end{equation}
Due to the convexity of the relative entropy,
\begin{equation}
S\left(\sum_i q_i \rho_i\Vert\sum_i q_i \sigma_i\right)
\le \sum_i q_i S(\rho_i||\sigma_i),
\end{equation}
we can further tighten the expression of the relative entropy by taking 
its convex hull. 
(Via the convexification process, i.e., the convex hull construction, the corresponding
separable state can also be obtained.)\thinspace\ 
Therefore, we arrive at an upper bound for the relative 
entropy of entanglement of the mixed state $\rho(\{p\})$:
\begin{equation}
\label{eqn:conjecture}
E_{\rm R}\left(\rho(\{p\})\right)\le {\rm co}F(\{p\}),
\end{equation}
where 
\begin{eqnarray}
\label{eqn:F}
F(\{p\})\equiv\sum_k p_k 
\log_2 \frac{p_k}{C_k^n \cos^{2k}\theta \sin^{2(n-k)}\theta}=\sum_k p_k \log_2 \frac{p_k\, n^n}{C_k^n \alpha^{k} (n-\alpha)^{n-k}},
\end{eqnarray}
where the angle $\theta$ satisfies Eq.~(\ref{eqn:theta}), 
$C_k^n\equiv n!/\big(k!(n-k)!\big)$, and $\alpha\equiv \sum_k p_k\, k$. 

Having established an upper bound for RE for the state $\rho(\{p\})$, 
we now make the restriction to mixtures of two distinct $n$-qubit states 
$\ket{S(n,k_1)}$ and $\ket{S(n,k_2)}$ (with $k_1\ne k_2$): 
\begin{eqnarray}
\rho_{n;k_1,k_2}(s)\equiv s\ketbra{S(n,k_1)} 
+(1-s)\ketbra{S(n,k_2)}.
\end{eqnarray}
One trivial example is $\rho_{n;0,n}(s)$, which is obviously unentangled
as it is the mixture of two separable pure states $\ket{0^{\otimes n}}$
and $\ket{1^{\otimes n}}$.  Other mixtures are generally entangled, except
possibly at the end-points $s=0$ or $s=1$ when the mixture contains either 
$\ket{S(n,0)}$ or $\ket{S(n,n)}$.  We first investigate the two-qubit (i.e.\ $n=2$) 
case.  Besides the trivial mixture, $\rho_{2;0,2}$, there is only one inequivalent
mixture, $\rho_{2;0,1}(s)$ [which is equivalent to $\rho_{2;2,1}(s)$], which
is---up to local basis change---the so-called {\it maximally entangled mixed 
state\/}~\cite{MunroEtAl,WeiEtAl} (for a certain range of $s$)
\begin{equation}
\rho_{2;0,1}=s\,\ketbra{11}+(1-s)\ketbra{\Psi^+},
\end{equation}
where $\ket{\Psi^+}\equiv(\ket{01}+{10})/\sqrt{2}$.
The function $F$ for this state [denoted by $F_{2;0,1}(s)$] is 
\begin{equation}
\label{eqn:rho201}
F_{2;0,1}(s)=s \,\log_2\frac{4s}{(1+s)^2}+(1-s)\log_2\frac{2}{1+s}\,,
\end{equation}
which is convex in $s$.  It is exactly the expression for the relative entropy 
of entanglement for the state $\rho_{2;0,1}$ found by Vedral and 
Plenio~\cite{VedralPlenio98} (see their Eq.~(56) with $\lambda$ replaced by $1-s$).

\begin{figure}
\centerline{\psfig{figure=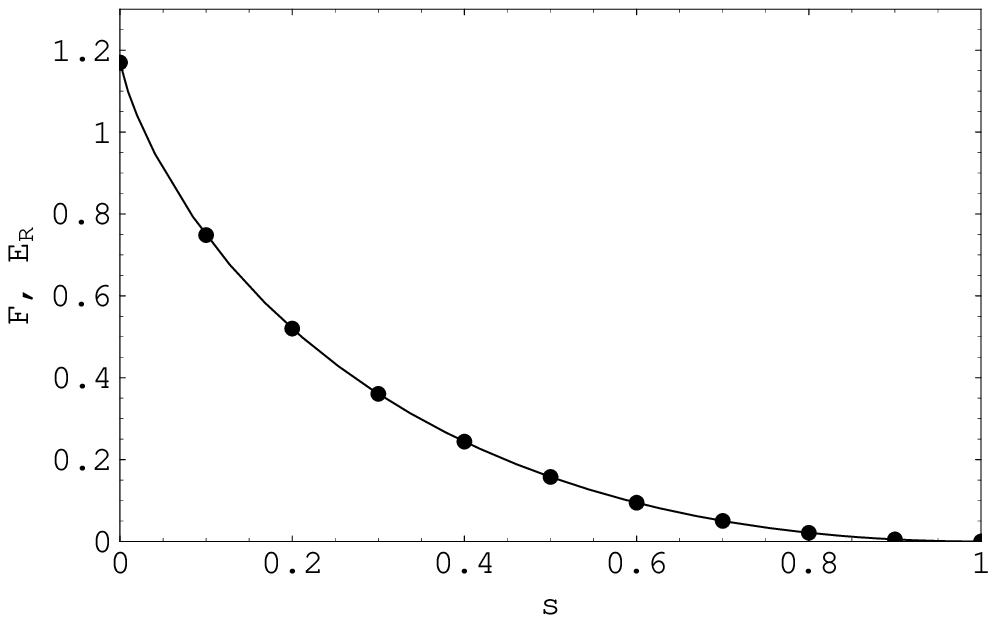,width=7cm,height=4cm,angle=0}}
\vspace{0.2cm}
\centerline{\psfig{figure=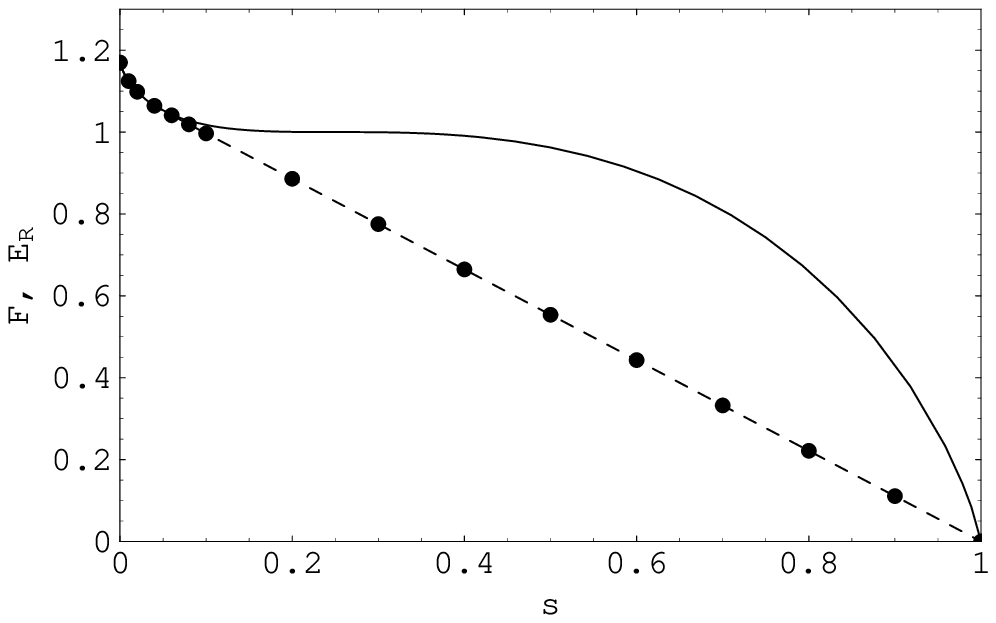,width=7cm,height=4cm,angle=0}}
\vspace{0.2cm}
\centerline{\psfig{figure=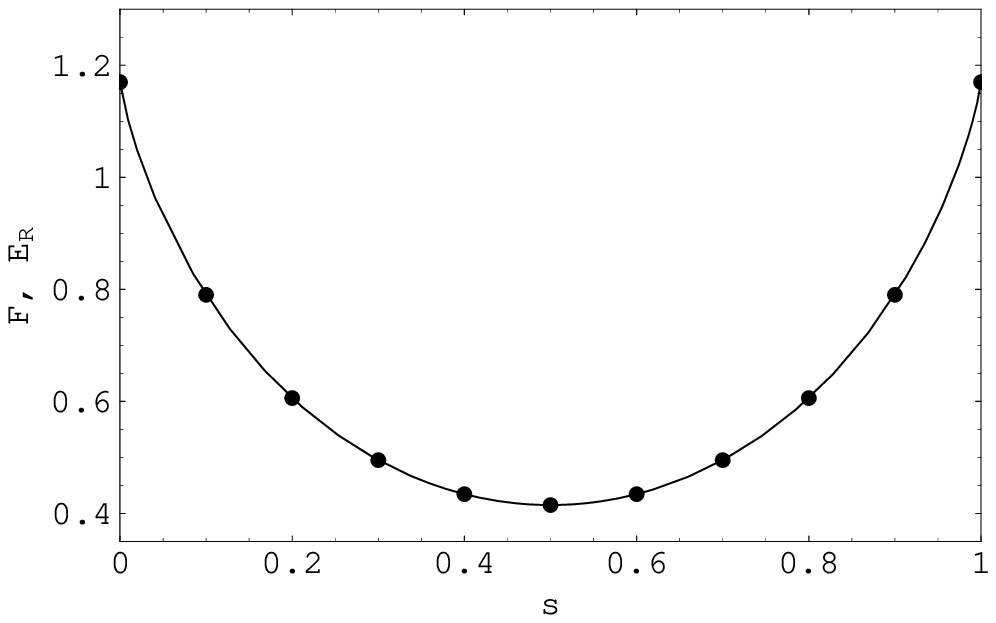,width=7cm,height=4cm,angle=0}}
\caption{Comparison of $F$ (solid curve), ${\rm co}\,F$ (convexification indicated
by dashed line) and the numerical value of $E_{\rm R}$ (dots) for the states
$\rho_{3;0,1}(s)$, $\rho_{3;0,2}(s)$, and $\rho_{3;1,2}(s)$ (from top to bottom).  
Note that the $\log$ function is implicitly base-2.
}
\label{fig:Er3}
\end{figure}

For $n=3$ there are three other inequivalent mixtures:
$\rho_{3;0,1}(s)$ 
[equivalent to $\rho_{3;3,2}(s)$], 
$\rho_{3;0,2}(s)$ 
[to $\rho_{3;3,1}(s)$], and 
$\rho_{3;1,2}(s)$ 
[to $\rho_{3;2,1}(s)$]. 
In Fig.~\ref{fig:Er3} we compare 
the function $F$ in Eq.~(\ref{eqn:F}), 
its convex hull  ${\rm co}\,F$, 
and numerical values of $E_{\rm R}$ 
obtained using the general scheme described in 
Ref.~\cite{VedralPlenio98} extended beyond the two-qubit case. 
The agreement between ${\rm co}\,F$ and the numerical values 
of $E_{\rm R}$ appears to be exact.

\begin{figure}
\centerline{\psfig{figure=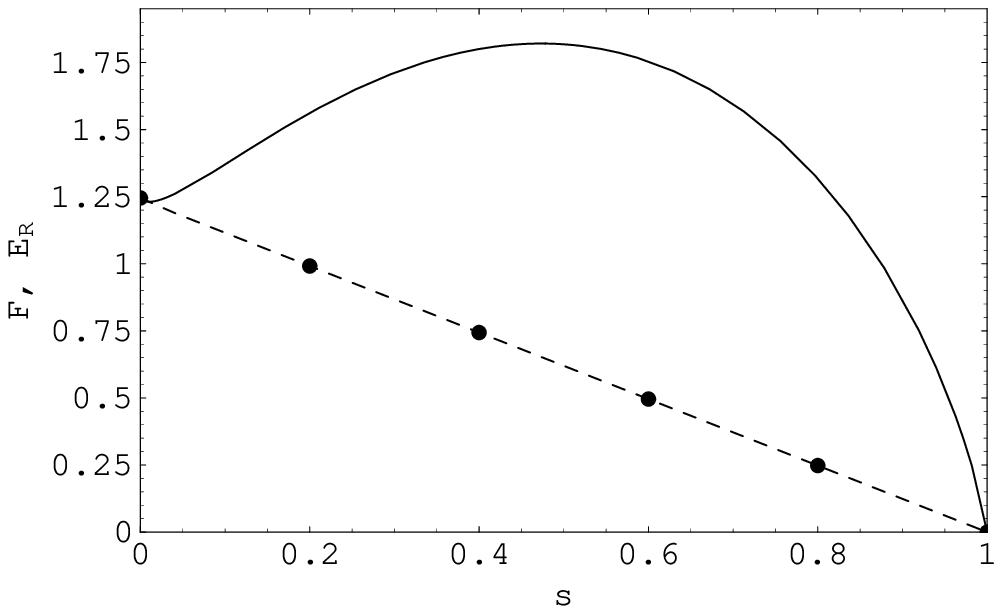,width=7cm,height=4cm,angle=0}}
\vspace{0.2cm}
\centerline{\psfig{figure=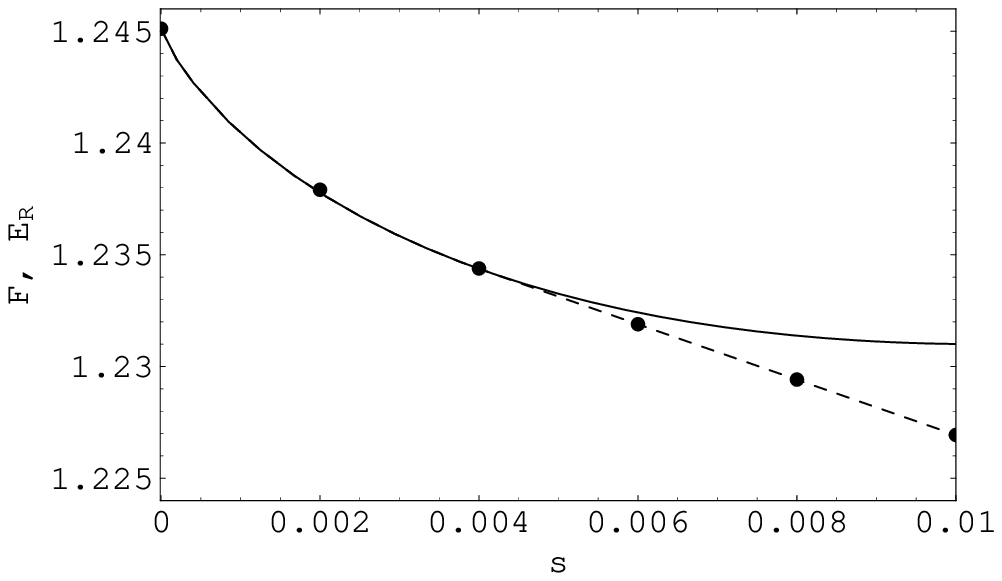,width=7.4cm,height=4cm,angle=0}}
\caption{Comparison of $F$, its convex hull, and the numerical value of $E_{\rm R}$ for the state
 $\rho_{4;0,3}(s)$. Upper panel shows the whole range $s\in[0,1]$,
 whereas the lower panel shows a blow-up of the range $s\in[0,0.01]$.  
}
\label{fig:Er4A}
\end{figure}

\begin{figure}
\centerline{\psfig{figure=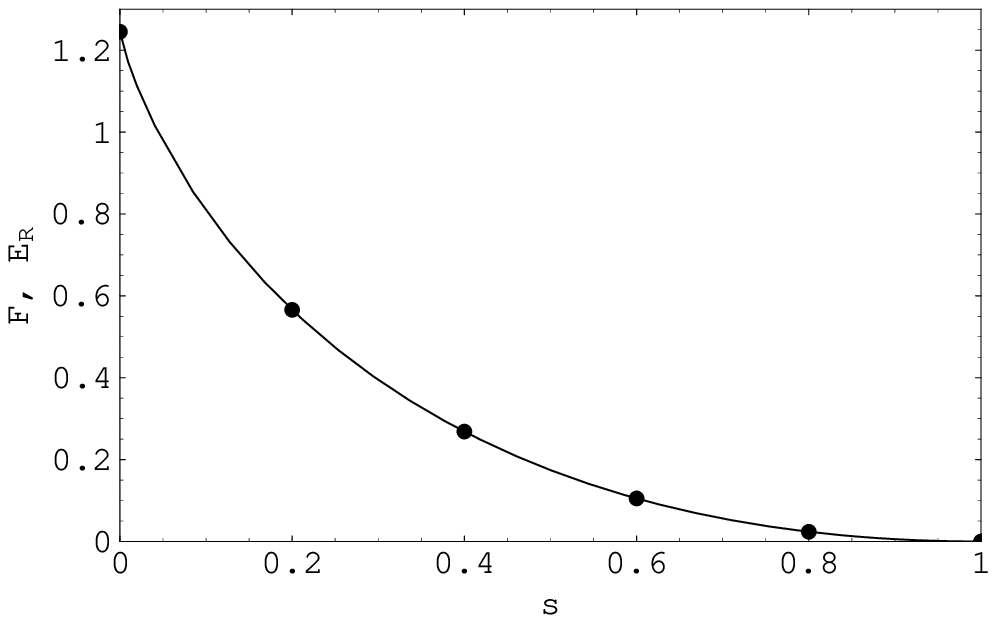,width=7cm,height=4cm,angle=0}}
\vspace{0.2cm}
\centerline{\psfig{figure=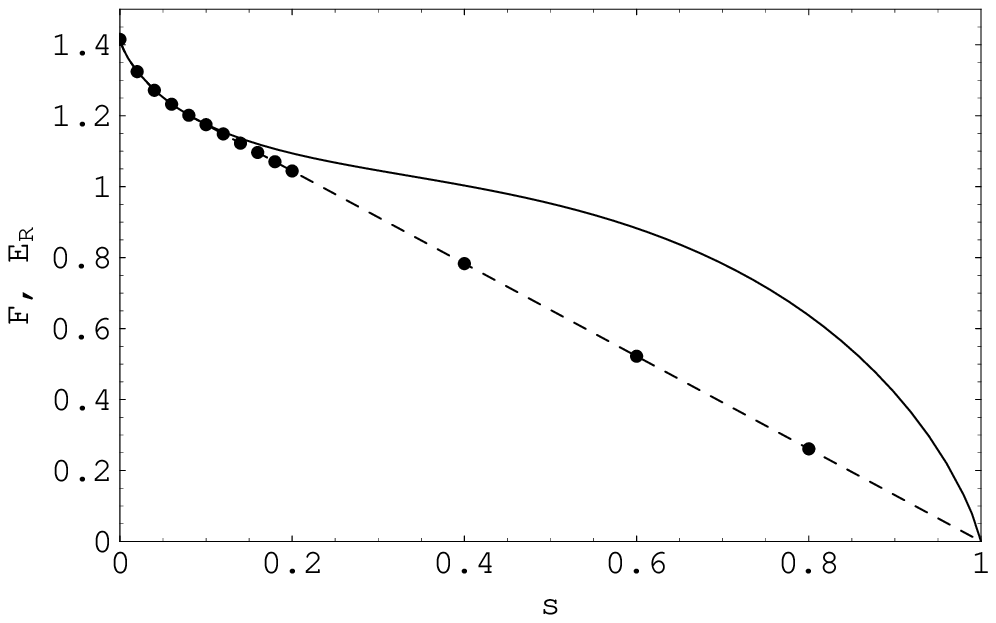,width=7cm,height=4cm,angle=0}}
\vspace{0.2cm}
\centerline{\psfig{figure=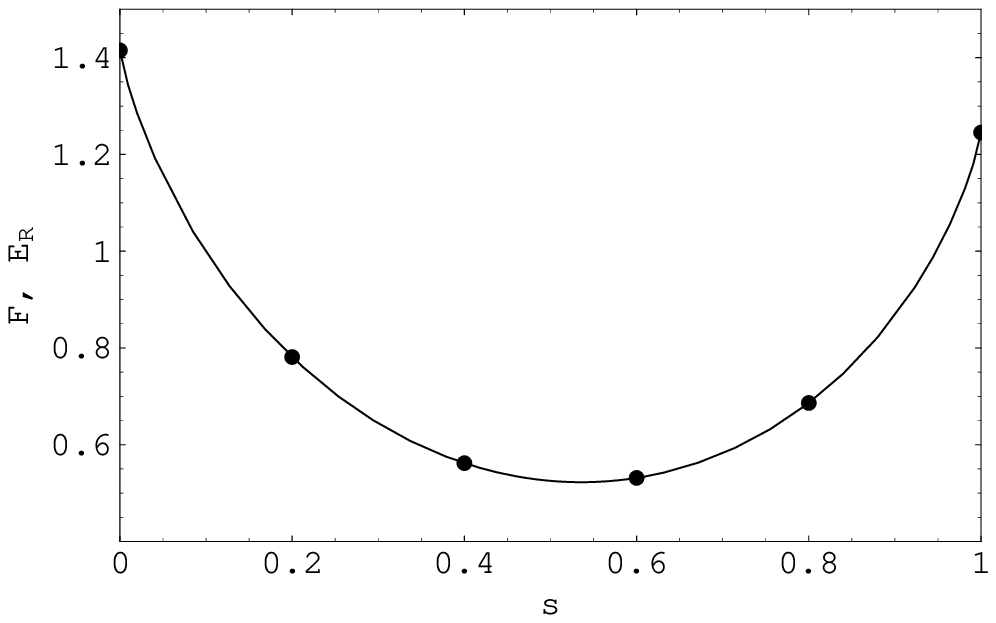,width=7cm,height=4cm,angle=0}}
\vspace{0.2cm}
\centerline{\psfig{figure=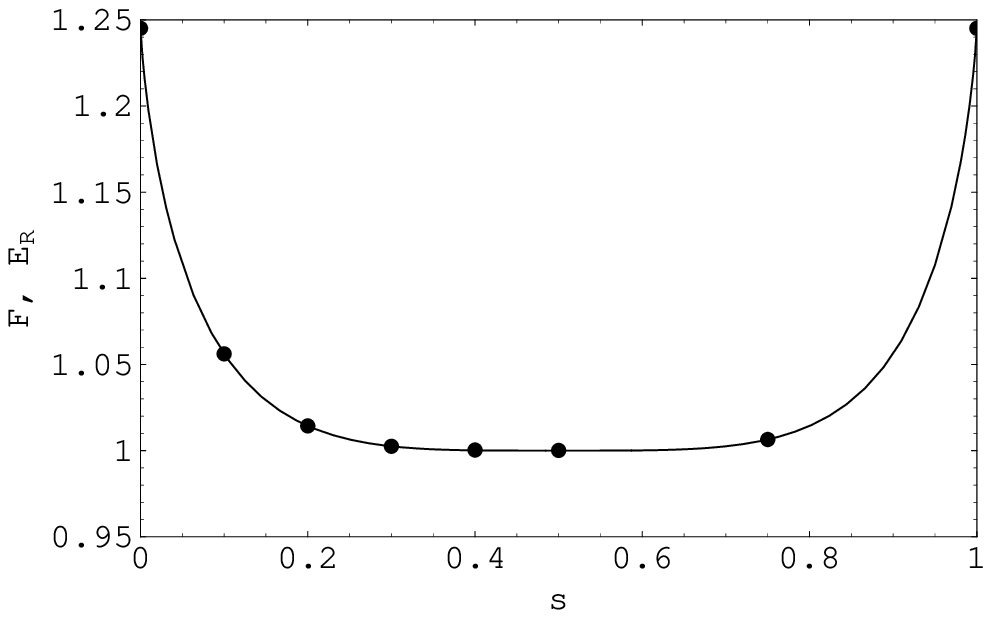,width=7.2cm,height=4cm,angle=0}}
\caption{Comparison of $F$, its convex hull, and the numerical value of $E_{\rm R}$ for the states
$\rho_{4;0,1}(s)$, $\rho_{4;0,2}(s)$, $\rho_{4;1,2}(s)$, and $\rho_{4;1,3}(s)$ (from top to bottom).  
}
\label{fig:Er4B}
\end{figure}

For $n=4$ there are five inequivalent nontrivial mixtures:
$\rho_{4;0,1}(s)$, 
$\rho_{4;0,2}(s)$, 
$\rho_{4;0,3}(s)$, 
$\rho_{4;1,2}(s)$, and 
$\rho_{4;1,3}(s)$.  
In Figs.~\ref{fig:Er4A} and \ref{fig:Er4B} we again compare 
the function $F$ in Eq.~(\ref{eqn:F}), 
its convex hull ${\rm co}\,F$, 
and numerical values of $E_{\rm R}$. 
Again the agreement between ${\rm co}\,F$ and the numerical values 
of $E_{\rm R}$ appears to be exact.

From these agreements, we are led to the following conjecture:
\\
\noindent
{\it Conjecture}~1: 
The relative entropy of entanglement $E_{\rm R}\left(\rho(\{p\})\right)$ 
for the mixed states $\rho(\{p\})$ is given exactly by ${\rm co}F(\{p\})$.

\begin{figure}
\centerline{\psfig{figure=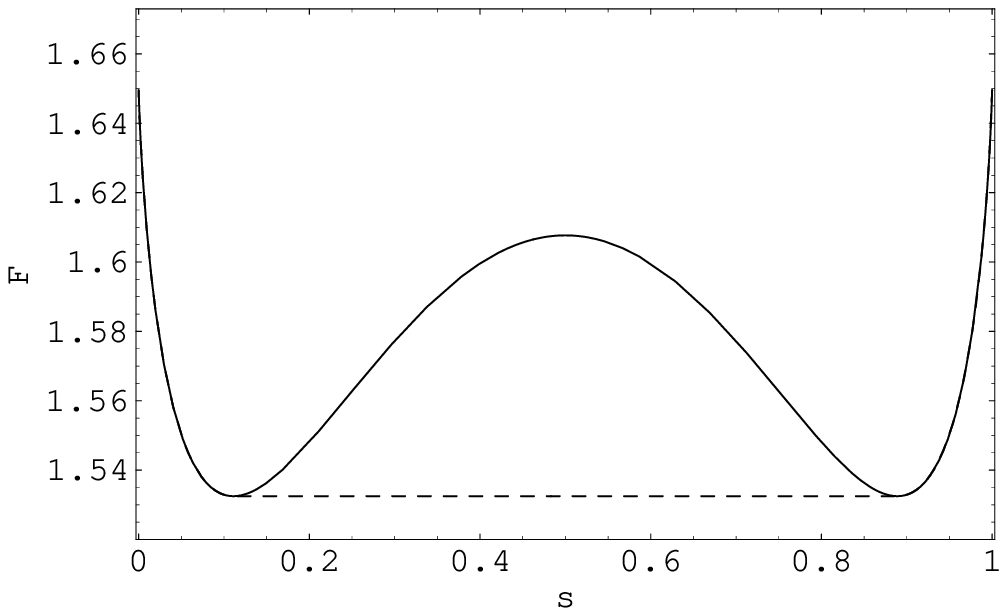,width=7cm,height=4.5cm,angle=0}}
\caption{The function $F$ (solid curve) and its convex hull (dashed line indicates convexification) for  the seven-qubit mixed state
$\rho_{7;2,5}(s)$. 
}
\label{fig:Er725}
\end{figure}
For the states that we have just considered, we now pause to give the formulas 
for $E_{\rm R}$ suggested by the conjecture.  For the three-qubit mixed state 
$\rho_{3;2,1}(s)$, its conjectured $E_{\rm R}$ is
\begin{subequations}
\begin{equation}
\label{eqn:ErWW}
s\log_2\frac{9s}{(1+s)^2(2-s)}+
(1-s)\log_2\frac{9(1-s)}{(2-s)^2(1+s)}.
\end{equation}
For $\rho_{3;0,1}(s)$, it is 
\begin{equation}
\label{eqn:rho301}
s\log_2\frac{27s}{(2+s)^3}+
(1-s)\log_2\frac{9}{(2+s)^2}.
\end{equation}
\end{subequations}
For $\rho_{4;0,1}(s)$, it is 
\begin{subequations}
\begin{equation}
s\log_2\frac{256s}{(3+s)^4}+
(1-s)\log_2\frac{64}{(3+s)^3}. 
\end{equation}
For $\rho_{4;1,2}(s)$, it is 
\begin{equation}
s\log_2\frac{64s}{(2\!-\!s)(2\!+\!s)^3}+
(1\!-\!s)\log_2\frac{128(1-s)}{3(2\!-\!s)^2(2\!+\!s)^2}.
\end{equation}
For $\rho_{4;1,3}(s)$, it is 
\begin{equation}
s\log_2\frac{64s}{(3\!-\!2s)(1\!+\!2s)^3}+
(1\!-\!s)\log_2\frac{64(1-s)}{(3\!-\!2s)^3(1\!+\!2s)}.
\end{equation}
\end{subequations}
For states such as $\rho_{3;0,2}$, $\rho_{4;0,2}$, and $\rho_{4;0,3}$,
convexifications (i.e. convex hull constructions) are needed; see Figs.~\ref{fig:Er3}, \ref{fig:Er4A}, and \ref{fig:Er4B}.
In Fig.~\ref{fig:Er725} we give an example of a seven-qubit state, viz., 
$\rho_{7;2,5}(s)$.

Although we have not been able to prove our conjecture, we have observed some 
supporting evidence, in addition to the numerical evidence presented above. 
We begin by noting that the states $\rho(\{p\})$ are invariant under the 
projection 
\begin{equation}
\label{eqn:Projection}
{\bf \rm P}:\rho\rightarrow 
\int\frac{d\phi}{2\pi}\, 
U(\phi)^{\otimes n}\rho\, U(\phi)^{\dagger\otimes n}
\end{equation}
with $U(\phi)\big\{\ket{0},\ket{1}\big\}\to
\big\{\ket{0},{\rm e}^{-i\phi}\ket{1}\big\}$. 
Vollbrecht and Werner~\cite{VollbrechtWerner01}
have shown that in order to find the closest separable mixed state
for a state that is invariant under projections such as ${\bf \rm P}$,
it is only necessary to search within the separable states that are
also invariant under the projection.  We can further reduce the
set of separable states to be searched by invoking another symmetry property
possessed by $\rho(\{p\})$: these states are also, by construction,
invariant under permutations of all parties.  Let us denote by $\Pi_i$ one 
of the permutations of parties, and by $\Pi_i(\rho)$ the state obtained
from $\rho$ by permuting the parties under $\Pi_i$.  We now show that 
the set of separable states to be searched can be reduced to the 
separable states that are invariant under the permutations.  
To see this, suppose that $\rho$ is a mixed state in the family~(\ref{eqn:mixSnk}), 
and that $\sigma^*$ is one of the closest separable states to $\rho$, i.e.,
\begin{equation}
E_{\rm R}(\rho)\equiv\min_{\sigma\in {\cal D}} S(\rho||\sigma)=S(\rho||\sigma^*).
\label{eqn:extreme}
\end{equation}
As $\rho$ is invariant under all $\Pi_i$, we have
\begin{equation}
E_{\rm R}(\rho)=\frac{1}{N_\Pi}\sum_i S\left(\rho\big\Vert\Pi_i(\sigma^*)\right),
\end{equation}
where $N_\Pi$ is the number of permutations.
By using the convexity of the relative entropy we have
\begin{equation}
E_{\rm R}(\rho)\ge S\left(\rho\big\Vert\big[\sum_i \Pi_i(\sigma^*)/N_\Pi\big]\right).
\end{equation}
However, because of the extremal property, Eq.~(\ref{eqn:extreme}),
the inequality must be saturated, as the left-hand side is already minimal.  
This shows that 
\begin{equation}
\sigma^{**}\equiv \frac{1}{N_\Pi}\sum_i \Pi_i(\sigma^*)
\end{equation}
also a closest separable mixed state to $\rho$, and is manifestly invariant under all 
permutations.  Thus, we only need to search within this restricted family of separable states.

It is not difficult to see that the set ${\cal D}_S$ of all separable mixed states 
that are diagonal in the basis of $\{\ket{S(n,k)}\}$ can be constructed from
a convex mixture of separable states in Eq.~(\ref{eqn:sigmatheta}).
That is, for any $\sigma_s\in {\cal D}_S$ we have a decomposition 
\begin{equation}
\label{eqn:sigmas}
\sigma_s=\sum_i t_i \,\sigma(\theta_i),
\end{equation}
where $t_i\ge 0$, $\sum_i t_i=1$, and $\sigma(\theta_i)$ is of the form~(\ref{eqn:sigmatheta}).
This is because the separability of the states~(\ref{eqn:mixSnk}) implies
that there exists a decomposition into pure states such that
each pure state is a separable state. Furthermore,
 because $\{\ket{S(n,k)}\}$ are eigenstates of $\rho(\{p\})$,
the most general
form of the pure state in its decomposition is 
\begin{equation}
\sum_k\sqrt{q_k}\,e^{i\phi_k}\ket{S(n,k)}.
\end{equation}
This pure state is separable if and only if it
is of the form~(\ref{eqn:xi}), up to an overall irrelevant phase.  
As $\rho(\{p\})$ is invariant under
the projection ${\rm P}$~(\ref{eqn:Projection}), a pure state in Eq.~(\ref{eqn:xi})
will be projected to the mixed state in Eq.~(\ref{eqn:sigmatheta}) under
${\rm P}$. Thus, every separable state that is diagonal in $\{\ket{S(n,k)}\}$
basis can be expressed in the form~(\ref{eqn:sigmas}).

\begin{figure}
\centerline{\psfig{figure=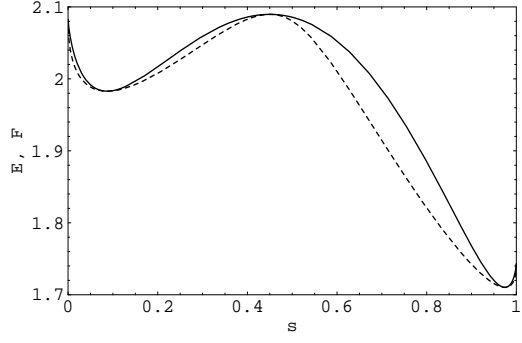,width=7cm,height=4.5cm,angle=0}}
\caption{Comparision of ${\cal E}$ (dashed curve) and $F$ (solid curve) for the eleven-qubit mixed state
$\rho_{11;2,6}(s)$.  
}
\label{fig:ElogEr1126}
\end{figure}
Hence, our conjecture~(\ref{eqn:conjecture})  ensures (via any necessary convexification) that it is at least the minimum (of the relative entropy) when the separable mixed states are restricted to ${\cal D}_S$. 
However, in order to prove the conjecture, one would still need to show that 
the expression is also the minimum when the restirction to ${\cal D}_S$
is relaxed. 

We remark that our conjecture is consistent with the results of 
Ishizaka~\cite{Ishizaka02}, in that our conjectured $\sigma^*$ 
satisfies the condition that $[\rho,\sigma^*]=0$ and that $\sigma^*$ 
has the same reduction as $\rho$ for every party. 
{
Furthermore, suppose
$\sigma^*$ (diagonal in the basis $\{\ket{S(n,k)}\}$) represents the separable state that gives the conjectured value of RE:
\begin{equation}
\sigma^*=\sum_k r_k \ket{S(n,k)}\bra{S(n,k)},
\end{equation} 
where the $r$'s can be obtained by finding the convex hull of the function
$F$ in Eq.~(\ref{eqn:F}). Now consider any separable state $\tau$ in
the Hilbert space {\it orthogonal\/} to the subspace spanned by $\{S(n,k)\}$.
We need to show that the separable state $\sigma(x)\equiv x\sigma^*+(1-x) \tau$,
for any $x\in [0,1]$, gives greater relative entropy with $\rho(\{p\})$ in Eq.~(\ref{eqn:mixSnk}) than 
$\sigma^*$ does with $\rho(\{p\})$, i.e., 
\begin{equation}
\label{eqn:Sinequality}
S\left(\rho(\{p\})\Vert\sigma(x)\right)\ge S\left(\rho(\{p\})\Vert\sigma^*\right).
\end{equation}
Writing out the expression explicitly, we have that
\begin{eqnarray}
S\left(\rho(\{p\})\Vert\sigma(x)\right)=\sum_k p_k \log\frac{p_k}{x\,r_k} 
\ge\sum_k p_k \log\frac{p_k}{ r_k}= S\left(\rho(\{p\})\Vert\sigma^*\right).
\end{eqnarray}
Note that $\tau$ gives no contribution in the relative entropy, as
it is orthogonal to $\rho(\{p\})$, and that we have not used the
fact that $\tau$ is separable. But to prove Conjecture 1 we need
to show that Eq.~(\ref{eqn:Sinequality}) holds if separable $\tau$
is not orthogonal to the subspace spanned by $\{S(n,k)\}$.
}

Recall that for pure states we found the inequality $E_{\log_2}\le E_{\rm R}$. 
Does this inequality hold for mixed states? We do not know the complete answer to this 
question, but for the mixed state $\rho(\{p\})$ we shall at least find that this inequality 
would hold if Conjecture~1 holds.  To see this, we first establish that ${\cal E}(\{q\})$
is a lower bound on $F(\{q\})$; see the example in Fig.~\ref{fig:ElogEr1126}. 
The proof is as follows. Recall that 
\begin{subequations}
\begin{equation}
{\cal E}(\{p\})=
-2\log_2\left[\max_\theta \sum_k \sqrt{p_k}\, \sqrt{C_k^n} \cos^k\theta\sin^{n-k}\theta\right].
\end{equation}
By the concavity of $\log$, we then have
\begin{eqnarray}
-2\log_2\left[\sum_k \sqrt{p_k}\, \sqrt{C_k^n} \cos^k\theta\sin^{n-k}\theta\right] 
\le \sum_k p_k \log_2\frac{p_k}{C_k^n \cos^{2k}\theta \sin^{2(n-k)}\theta}.
\end{eqnarray}
Hence
\begin{eqnarray}
\min_\theta -2\log_2\left[\sum_k \sqrt{p_k}\, \sqrt{C_k^n} \cos^k\theta\sin^{n-k}\theta\right] 
\le \min_\theta\sum_k p_k \log_2\frac{p_k}{C_k^n \cos^{2k}\theta \sin^{2(n-k)}\theta},
\end{eqnarray}
or equivalently
\begin{equation}
{\cal E}(\{p\})\le F(\{p\}).
\end{equation}
If Conjecture~1 is correct then by taking the convex hull of both sides of this inequality we would have
\begin{equation}
E_{\log_2}\le E_{\rm R}
\end{equation}
for the family of states~(\ref{eqn:mixSnk}).
\end{subequations}
Notice that we have also shown that this relation holds for arbitrary pure states.  
It would be interesting to know whether it also holds for arbitrary mixed states.

\section{Concluding remarks}
\noindent
\label{sec:summary}
We have provided a lower bound on the relative entropy of entanglement
for arbitrary multipartite pure states in terms of their geometric
measure of entanglement.  For several families of pure states 
we have shown that the bound is in fact saturated, and thus provides 
the exact value of the relative entropy of entanglement. For mixtures of certain
permutation-invariant states we have conjectured analytic expressions for the
relative entropy of entanglement.

It is possible that our results on the relative entropy of entanglement might be 
applicable to the checking of the consistency of some equalities and inequalities~\cite{PlenioVedral01,WuZhang00,GalvaoPlenioVirmani00} 
regarding minimal reversible entanglement generating sets (MREGSs).
Consider, e.g., the particular family of $n$-qubit pure states $\{\ket{S(n,k)}\}$, 
the relative entropy of entanglement of which we have given in Eq.~(\ref{eqn:rhoSnk}).  
Now, if we trace over one party we get a mixed $(n-1)$-qubit state:
\begin{equation}
\label{eqn:Tr1}
{\rm Tr}_1 \ketbra{S(n,k)}=\frac{n\!-\!k}{n}\ketbra{S(n\!-\!1,k)}+\frac{k}{n}\ketbra{S(n\!-\!1,k\!-\!1)}.
\end{equation} 
We have also given a conjecture for the relative entropy of entanglement for this 
mixed state.  If we trace over $m$ parties, the reduced mixed state would be a mixture
of $\{\ket{S(n-m,q)}\}$ [with $q\le (n-m)$], and again we have given a conjecture for 
its relative entropy of entanglement.  For example, if we start with $\ket{S(4,1)}$, and
trace over one party and then another, we get the sequence:
\begin{equation}
\ket{S(4,1)}\rightarrow \rho_{3;0,1}({1}/{4})
\rightarrow
\rho_{2;0,1}({1}/{2}),
\end{equation}
for which we have given the corresponding relative entropies of entanglement 
in Eqs.~(\ref{eqn:rhoSnk}), (\ref{eqn:rho301}) and (\ref{eqn:rho201}). 
(To be precise, the second formula is a conjecture; the others are proven.)\thinspace\ 
The afore-mentioned equalities and inequalities concerning MREGS usually involve
only the von Neumann entropy and the regularized (i.e.~asymptotic) relative entropy of 
entanglement of the pure state and its reduced density matrices.
The regularized relative entropy of entanglement is defined as
\begin{equation}
E_{\rm R}^\infty(\rho)\equiv
\lim_{n\rightarrow \infty}\frac{1}{n}E_{\rm R}(\rho^{\otimes n}).
\end{equation}
The calculation of the regularized relative entropy of entanglement is, in general, 
much more difficult than for the non-regularized case, and the (in)equalities 
involving the regularized relative entropy of entanglement are thus difficult to 
check.  Nevertheless, it is known that $E_{\rm R}^\infty\le E_{\rm R}$, so we can
check their weaker forms by replacing $E_{\rm R}^\infty$ by $E_{\rm R}$, and the 
corresponding (in)equalities by weaker inequalities.   

{ 
Plenio and Vedral~\cite{PlenioVedral01} have derived a lower bound on the RE of a tripartite pure state $\rho_{\rm ABC}=\ketbra{\psi}$ in terms of the the entropies and
RE's of the reduced states of two parties:
\begin{equation}
\max\{E_{\rm R}(\rho_{\rm AB})+S(\rho_{\rm AB}),E_{\rm R}(\rho_{\rm AC})+S(\rho_{\rm AC}),E_{\rm R}(\rho_{\rm BC})+S(\rho_{\rm BC})\}\le E_{\rm R}(\rho_{\rm ABC}),
\end{equation}
where $\rho_{\rm AB}={\rm Tr}_C(\rho_{\rm ABC})$ (and similarly for $\rho_{\rm AC}$
and $\rho_{\rm BC}$) and $S(\rho)\equiv - {\rm Tr}\rho\log_2\rho$ is
the von Neumann entropy. They have further found that this lower bound
is saturated by $\ket{\rm GHZ}$ and $\ket{\rm W}$. This raises an
interesting question~\cite{saturation}: is the above lower bound (for
$n$-partite pure states) saturated by the states that saturate
the lower bound $E_{\log_2}=-2\log_2{\Lambda_{\max}}(\psi)\le E_{\rm R}(\psi)$? Numerical tests 
seem to suggest that the Plenio-Vedral bound is tighter than
$E_{\log_2}$. If this is the case then all states that saturate
the lower bound $E_{\log_2}$ on $E_{\rm R}$ will saturate the
Plenio-Vedral bound. Based on Conjecture~1, we can show that for
$\rho_{12\ldots n}=\ketbra{S(n,k)}$ the inequality
\begin{equation}
\label{eqn:CRE}
\max_i \{E_{\rm R}(\rho_{12\ldots\hat{i}\ldots n})+S(\rho_{ 12\ldots\hat{i}\ldots n})\}\le E_{\rm R}(\rho_{12\ldots n})
\end{equation}
is saturated, where $\rho_{12\ldots\hat{i}\ldots n}\equiv{\rm Tr}_i(\rho_{12\ldots n})$ is the reduced density matrix obtained
from $\rho_{12\ldots n}$ by tracing out the $i$-th party. The proof
is as follows. As $\ket{S(n,k)}$
is permutation-invariant, there is no need to maximize over all parties,
and we can simply take $i=1$, obtaining the reduced state $\rho_{n-1;k-1,k}(k/n)$ as
in Eq.~(\ref{eqn:Tr1}). As the corresponding function $F_{n-1;k-1,k}(s)$
of $\rho_{n-1;k-1,k}(s)$ is convex for $s\in[0,1]$, we immediately obtain from Conjecture~1 that, for $\rho_{n-1;k-1,k}(k/n)$,
\begin{subequations}
\begin{eqnarray}
\!\!\!\!\!\!\!\!\!\!\!E_{\rm R}\left(\rho_{n\!-\!1;k\!-\!1,k}(k/n)\right)&=&\log_2\left[C^n_k \left(\frac{k}{n}\right)^k\left(\frac{n\!-\!k}{n}\right)^{n\!-\!k}\right]
+\frac{k}{n}\log_2\frac{k}{n}+\frac{n\!-\!k}{n}\log_2\frac{n\!-\!k}{n}\\
&=&E_{\rm R}\left(\ket{S(n,k)}\right)- S\left(\rho_{n\!-\!1;k\!-\!1,k}(k/n)\right).
\end{eqnarray}
\end{subequations}
Therefore, the bound in Eq.~(\ref{eqn:CRE}) is saturated for $\rho_{12\ldots n}=\ketbra{S(n,k)}$.
}
 
A major challenge is to extend the ideas contained in the present Paper from 
the relative entropy of entanglement to its regularized version, the latter in fact being of wider interest than the former.
{The alternative way of defining
the relative entropy via the optimization over PPT states may also been
used, in view of the recent progress on the bipartite regularized relative
entropy of entanglement~\cite{AudenaertEtAl}. 
}

We now explore the possibility that the geometric measures can provide lower bounds 
on yet another entanglement measure---the entanglement of formation. 
If the relationship $ E_{\rm R} \le E_{\rm F}$ between the two measures of entanglement---the 
relative entropy of entanglement $E_{\rm R}$ and the entanglement of formation $E_{\rm F}$---should 
continue to hold for {\it multipartite\/} states (at least for pure states), 
and if $E_{\rm F}$ should remain a convex hull construction for mixed states, 
then we would be able to construct a lower bound on the entanglement of formation: 
\begin{eqnarray}
E_{\log_2}(\rho)& \equiv&\min_{p_i,\psi_i} \sum_i p_i E_{\log_2}(\ket{\psi_i})
\le \min_{p_i,\psi_i} \sum_i p_i E_{\rm R}(\ket{\psi_i}) \nonumber \\
&\le& \min_{p_i,\psi_i} \sum_i p_i E_{\rm F}(\ket{\psi_i})\equiv E_{\rm F}(\rho),
\end{eqnarray}
where $\{p_i\}$ and $\{\psi_i\}$ are such that $\rho=\sum_i p_i \ketbra{\psi_i}$.
Thus, $E_{\log_2}(\rho)$ is a lower bound on $E_{\rm F}(\rho)$.
By using the inequality $(1-x^2)\log_2e\le -2\log_2 x$  (for $0\le x\le 1$),
one further has has that 
$(\log_2e)E_{\sin^2}(\rho)\le  {E}_{\log_2}(\rho)\le E_{\rm F}(\rho)$.

\begin{figure}
\centerline{\psfig{figure=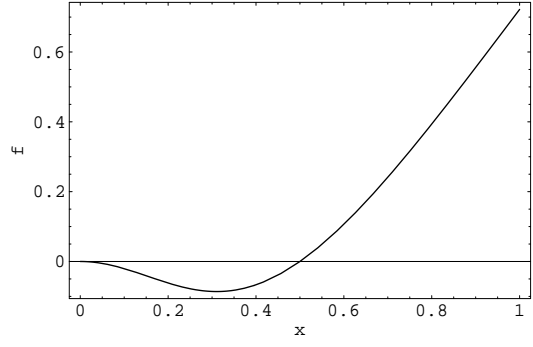,width=7cm,height=4.5cm,angle=0}}
\caption{ The function $f(4,x)$. It shows the violation of monotone condition~(\ref{eqn:violatemono})
when the function is negative.
}
\label{fig:violate}
\end{figure}
We remark that $E_{\sin^2}$ has been shown to be an entanglement
monotone~\cite{BarnumLinden01,WeiGoldbart03}, i.e., it is not increasing 
under local operations and classical communication (LOCC).  
However, $E_{\log_2}$ is {\it not\/} a monotone, as the following example shows.  
Consider the bipartite pure state
\begin{equation}
\ket{\psi}\equiv\frac{1}{\sqrt{1+N x^2}}\ket{00}+
\frac{x}{\sqrt{1+Nx^2}}\big(\ket{11}+\ket{22}+\dots+\ket{NN}\big), 
\end{equation}
with $|x|\le 1$, for which $E_{\log_{2}}=\log_{2}(1+N x^2)$.  
Suppose that one party makes the following measurement: 
\begin{equation}
{\cal M}_1\equiv\ket{0}\bra{0}, \ \ \ 
{\cal M}_2\equiv\ket{1}\bra{1}+\ket{2}\bra{2}+\dots+\ket{N}\bra{N}.
\end{equation}
With probability $P_1=1/(1+N x^2)$ the output state becomes $\ket{\psi_1}=\ket{00}$; 
with probability $P_2=N x^2/(1+N x^2)$ the output state becomes $\ket{\psi_2}=\big(\ket{11}+\ket{22}+\dots+\ket{NN}\big)/\sqrt{N}$, 
for which $E_{\log_{2}}=\log_{2}N$.  
For $E_{\log_{2}}$ to be a monotone it would be necessary that 
\begin{equation}
E_{\log_{2}}(\psi)\ge 
P_1 E_{\log_{2}}(\psi_1)+
P_2 E_{\log_{2}}(\psi_2).
\end{equation}
Putting in the corresponding values for the $P$'s and $E_{\log_{2}}$'s, 
we find that this inequality is  equivalent to
\begin{equation}
\label{eqn:violatemono}
f(N,x)\equiv\log_2(1+N x^2)- \frac{N x^2}{1+N x^2}\log_2 N\ge 0.
\end{equation}
As this is violated for certain values of $x$ with $N>2$, 
as exemplified in Fig.~\ref{fig:violate} for the plot of $f(4,x)$, 
we arrive at the conclusion that $E_{\log_{2}}$ is, in general, not a monotone.

\smallskip
\noindent
{\it Note added.\/} Certain results reported in the present Paper have recently been applied
by Vedral~\cite{Vedral04} to the macroscopic entanglement of $\eta$-paired superconductivity. 

\section*{Acknowledgments}
\noindent
We thank Vlatko Vedral and Pawel Horodecki for many useful discussions.
This work was supported by
NSF EIA01-21568 and
DOE DEFG02-91ER45439, as well as  
a Harry G.~Drickamer
Graduate Fellowship. M.E. acknowledges support from Wenner-gren Foundations.


\begin{thebibliography}{000} 
\bibitem{NielsenChuang00}
See, e.g., 
M. Nielsen and I. Chuang,
{\sl Quantum Computation and Quantum Information\/}
(Cambridge University Press, Cambridge, 2000).
\bibitem{Horodecki01}
For a review, see 
M. Horodecki, 
Quantum Inf. Comput. {\bf 1\/}, 3 (2001), 
and references therein.
\bibitem{BennettPopescuRohrlichSmolinThapliyal}
C. H. Bennett, S. Popescu, D. Rohrlich, J. A. Smolin, and A. V. Thapliyal,
Phys. Rev. A {\bf 63}, 012307 (2001).
\bibitem{BennettBernsteinPopescuSchumacher96}
C. H. Bennett, H. J. Berstein, S. Popescu, and B. Schumacher,
Phys. Rev. A {\bf 53}, 2046 (1996).
\bibitem{BennettDiVincenzoSmolinWootters96}
C. H. Bennett, D. DiVincenzo, J. Smolin, and W. K. Wootters,
Phys. Rev. A {\bf 53}, 3824 (1996).
\bibitem{Wootters98}
W. K. Wootters,
Phys. Rev. Lett. {\bf 80}, 2245 (1998).
\bibitem{PlenioVedral01}
M. B. Plenio and V. Vedral, J. Phys. A {\bf 34}, 6997 (2001).
\bibitem{VedralPlenioRippinKnight97}
V. Vedral, M. B. Plenio, M. A. Rippin, and P. L. Knight,
Phys. Rev. Lett. {\bf 78}, 2275 (1997).
\bibitem{Peres95}
A. Peres, Phys. Lett. A {\bf 202}, 16 (1995).
\bibitem{Note}
As we shall show later, this state has $E_{\rm R}=\log_2(9/4)$,
exactly the value cited in Ref.~\cite{PlenioVedral01} from a
numerical result.
\bibitem{Shimony95}
A. Shimony, 
Ann. NY. Acad. Sci. {\bf 755\/}, 675 (1995).
\bibitem{BarnumLinden01}
H. Barnum and N. Linden,
J. Phys. A {\bf 34\/}, 6787 (2001).
\bibitem{WeiGoldbart03}
T.-C. Wei and P. M. Goldbart, Phys. Rev. A {\bf 68}, 042307 (2003). 
\bibitem{WeiAltepeterGoldbartMunro03}
T.-C. Wei, J. B. Altepeter, P. M. Goldbart, and W. J. Munro, to appear in
Phys. Rev. A {\bf 70} (2004); quant-ph/0308031.
\bibitem{Ishizaka03}
S. Ishizaka, Phys. Rev. A {\bf 67}, 060301 (2003).
\bibitem{AudenaertEtAl}
K. Audenaert, J. Eisert, E. Jan\'e, M. B. Plenio, S. Virmani, and B. De Moor,
 Phys. Rev. Lett. {\bf 87}, 217902 (2001); K. Audenaert, M. B. Plenio, and J. Eisert, 
 Phys. Rev. Lett. {\bf 90}, 027901 (2003). We thank an
 anonymous referee for pointing out the alternative in defining
 the relative entropy of entanglement, as well as the above references.
\bibitem{Bravyi02}
S. Bravyi, Phys. Rev. A {\bf 67}, 012313 (2003).
\bibitem{VidalEtAl}
A. Ac\'\i n, G. Vidal, and J. I. Cirac, Quantum Inf. and Comput. {\bf 3}, 55 (2003).
G. Vidal and J. I. Cirac, Phys. Rev. Lett. {\bf 86}, 5803 (2001). We thank
an anonymous referee for pointing out these references.
\bibitem{Stockton}
J. K. Stockton, J. M. Geremia, A. C. Doherty, and H. Mabuchi,
Phys. Rev. A {\bf 67}, 022112 (2003).
\bibitem{MunroEtAl}
W. J. Munro, D. F. V. James, A. G. White, and P. G. Kwiat, Phys. Rev. A {\bf 64}, 030302 (2001).
\bibitem{WeiEtAl}
T.-C. Wei, K. Nemoto, P. M. Goldbart, P. G. Kwiat, W. J. Munro, and F. Verstraete, Phys. Rev. A {\bf 67}, 022110 (2003).
\bibitem{VedralPlenio98}
V. Vedral and M. B. Plenio, Phys. Rev. A {\bf 57}, 1619 (1998).
\bibitem{VollbrechtWerner01}
K. G. H. Vollbrecht and R. F. Werner, 
Phys. Rev. A {\bf 64\/}, 062307 (2001).
\bibitem{Ishizaka02}
S. Ishizaka, J. Phys. A: Math. Gen., {\bf 35}, 8075 (2002).
\bibitem{WuZhang00}
S. Wu and Y. Zhang, Phys. Rev. A {\bf 63}, 012308 (2000).
\bibitem{GalvaoPlenioVirmani00}
E. F. Galvao, M. B. Plenio, and S. Virmani, J. Phys. A {\bf 33}, 8809 (2000).
\bibitem{saturation}
We thank an anonymous referee for raising this question.
\bibitem{Vedral04}
V. Vedral, quant-ph/0405102.
\end{thebibliography}
\end{document}